\pdfoutput=1
\documentclass[useAMS,usenatbib] {mnras}
\usepackage[table]{xcolor}
\usepackage{amssymb}
\usepackage{float}

\usepackage{verbatim}
\usepackage{graphicx}
\usepackage{subcaption}
\usepackage{scalefnt}
\usepackage{amsmath}
\usepackage{txfonts}

\usepackage[T1]{fontenc}
\usepackage{aecompl}
\usepackage{array}
\usepackage{makecell}
\newcolumntype{x}[1]{>{\centering\arraybackslash}p{#1}}
\usepackage{tikz}

\title[Clumpy galaxies seen in H$\alpha$]{Clumpy galaxies seen in H-alpha: inflated observed clump properties due to limited spatial resolution and sensitivity}
\author[Tamburello et al.]{Valentina Tamburello,$^{1,2}$\thanks{E-mail: vtambure@physik.uzh.ch}, Alireza Rahmati$^{1}$, Lucio Mayer,$^{1,2}$, Antonio Cava$^3$, \newauthor Miroslava Dessauges-Zavadsky$^3$ and Daniel Schaerer$^{3,4}$\\
$^{1}$Center for Theoretical Astrophysics and Cosmology, Institute for Computational Science, University of Zurich,\\
Winterthurerstrasse 190, CH-8057 Z{\"u}rich, Switzerland\\
$^{2}$Physik-Institut, University of Zurich, Winterthurerstrasse 190, CH-8057 Z{\"u}rich, Switzerland\\
$^{3}$Observatoire de Gen\`eve, Universit\'e de Gen\`eve, 51 Ch. des Maillettes, 1290 Versoix, Switzerland\\
$^{4}$CNRS, IRAP, 14 Avenue E. Belin, 31400 Toulouse, France}
\begin{document}
\pagerange{\pageref{firstpage}--\pageref{lastpage}} 
\maketitle
\label{firstpage}
\begin{abstract}
High-resolution simulations of star-forming massive galactic discs have shown that clumps form with a characteristic baryonic mass in the range $10^7-10^8 M_{\odot}$, with a small tail exceeding $10^9$~M$_{\odot}$ produced by clump-clump mergers. This is in contrast with the observed kpc-size clumps with masses up to $10^{10}$~M$_{\odot}$ in high-redshift star-forming galaxies. In this paper we show that the comparison between simulated and observed star-forming clumps is hindered by limited observational spatial resolution and sensitivity. We post-process high-resolution hydrodynamical simulations of clumpy discs using accurate radiative transfer to model the effect of ionizing radiation from young stars and to compute H$\alpha$ emission maps. By comparing the intrinsic clump size and mass distributions with those inferred from convolving the H$\alpha$ maps with different gaussian apertures, we mimick the typical resolution used in observations. We found that with 100~pc resolution, mock observations can recover the intrinsic clump radii and stellar masses, in agreement with those found by lensing observations. Instead, using a 1~kpc resolution smears out individual clumps, resulting in their apparent merging. This causes significant overestimations of the clump radii and, therefore, masses derived using methods that use their observed sizes. We show that limited sensitivity can also force observations to significantly overestimate the clump masses. We conclude that a significant fraction of giant clumps detected in the observations may result from artificially inflated radii and masses, and that $\approx 100$~pc spatial resolution is required to capture correctly the physical characteristics of star-forming clumps if they are coherent structures produced by disc fragmentation.
\end{abstract}
%
%
\section{Introduction}\label{Intro}
%
%
Galaxies in the redshift range $1-3$ appear different from their counterparts in the Local Universe. They have higher velocity dispersions, lower rotation to dispersion ratios and higher gas fractions. One of the most significant peculiarities of high-z galaxies is that they show irregular morphologies in their stellar component which are commonly referred to as ``clumps'' \citep{Elmegreen2005}. These regions are also characterised by high star formation rates and can have sizes up to $1$~kpc and masses higher than $10^9$~M$_{\odot}$, which is much larger
than the largest star forming complexes known at low redshifts, such as Giant Molecular Clouds (GMCs).
As they appear to be sites of enhanced star formation, clumpy structures are also identified in H$\alpha$ observations of high redshift galaxies \citep{Genzel2011, Wisnioski2012, Swinbank2012, Livermore2012, Livermore2015}. The prevailing explanation for the existence of such oversized star forming complexes
at high redshift is that they are physically coherent objects, resulting from the fragmentation of massive gas-rich galactic discs induced by gravitational instability \citep{Ceverino2010, Bournaud2010}, although substructure might be present at smaller scale \citep{Bournaud2016}. This is based on the evidence that galactic discs at high redshift are indeed massive and considerably more gas-rich than at low redshift \citep{Tacconi2010}, implying that their characteristic scale of fragmentation 
as given by the Toomre instability theory should also be larger than at low resdhift, where GMCs-sized objects are instead produced \citep{Dekel2013}.
Yet observations become increasingly more difficult at higher redshifts, since galaxies become intrinsically smaller and their surface brightness decreases with redshift as $(1 + z)^4$. As a result only central bright regions of galaxies or extreme star-forming galaxies can be studied observationally. HST imaging of 0.15 arcsec allows to reach a resolution of $1-1.5$~kpc at $z=1-2$, while ground-bases telescopes, even with adaptive optics where a resolution of 0.25 arcsec is achieved, have a resolution of $\sim 2.1$~kpc at $z=1.5$. These resolutions are very close to the typical observationally measured clump sizes.
The fact that observed clump sizes are close to the resolution limits of typical observations is potentially worrisome in light of the prevailing theoretical interpretation based on a fragmentation scale as such sizes could be overestimated. Indeed, recent observations using gravitational lensing technique  which allow us to reach much higher spatial resolutions ($\sim 100$~pc), found that clumps in lensed galaxies are much smaller \citep{Jones2010, Adamo2013,  Livermore2012, Livermore2015}. However, it should be noted that the galaxies probed using the lensing technique are typically at $z \sim 1$ and are less massive than the other studied clumpy galaxies which are at higher redshifts.

A question arises spontaneously: are also the intrinsic sizes of clumps in the massive discs at  $z \sim 2$ smaller than usually claimed? Are their estimated masses correct or are they also affected by resolution limitations? First quantitative hints are provided by the recent numerical work from \citet{Behrendt2015}, who used one of the highest resolution simulations of massive clumpy gas discs to date and found no giant clumps. Instead, they found clump-like structures produced by gravitational instability with typical masses of the order a few times $10^7 M_{\odot}$.
Mimicking observations and the resolution of the instruments, they showed that their small clumps (radius $\sim 35$~pc) are distributed in 
loosely bound clusters, and thus can appear as giant kpc-clumps, since small-scale substructure would not be detectable in observations due to beam smearing.
Remarkably, the kinematics of the clusters of clumps they obtained are in quite good agreement with observations of individual giant clumps in the $z \sim 2$ galaxies 
of \citet{Genzel2011}. A caveat of these simulations is that they lack a stellar disc component and do not include star formation and feedback, which can affect fragmentation significantly
\citep{Moody2014, Tamburello2015, Oklopcic2016, Mayer2016, Mandelker2016}. However, the characteristic mass scale of
individual clumps that they found is comparable to that seen in lower resolution simulations of clumpy disc that include a stellar disc and a rich inventory of sub-grid physical
processes \citep{Tamburello2015} which include star formation and stellar feedback. 
In this paper we use selected simulations from  \citet{Tamburello2015}  to test if observed giant clumps could be
the result of blending of smaller structures owing to insufficient resolution in the observations.
To demonstrate this, we post-process our simulations in order to produce H$\alpha$ mock intensity maps, using radiative transfer calculations, augmented with noise and PSF 
specifications of instruments and observations employed in the literature for studying clumpy galaxies.

The paper is organised as follows. Section~2 describes the initial conditions and design of the simulations, Section~3 includes the results, and Section~4 summarises the paper with our conclusions.
%
%

\section{Methods}\label{Section2}

\subsection{Hydrodynamical Simulations}
%
%
\begin{figure*}
  \includegraphics[trim={1cm 0 0 0}, width=0.33\textwidth]{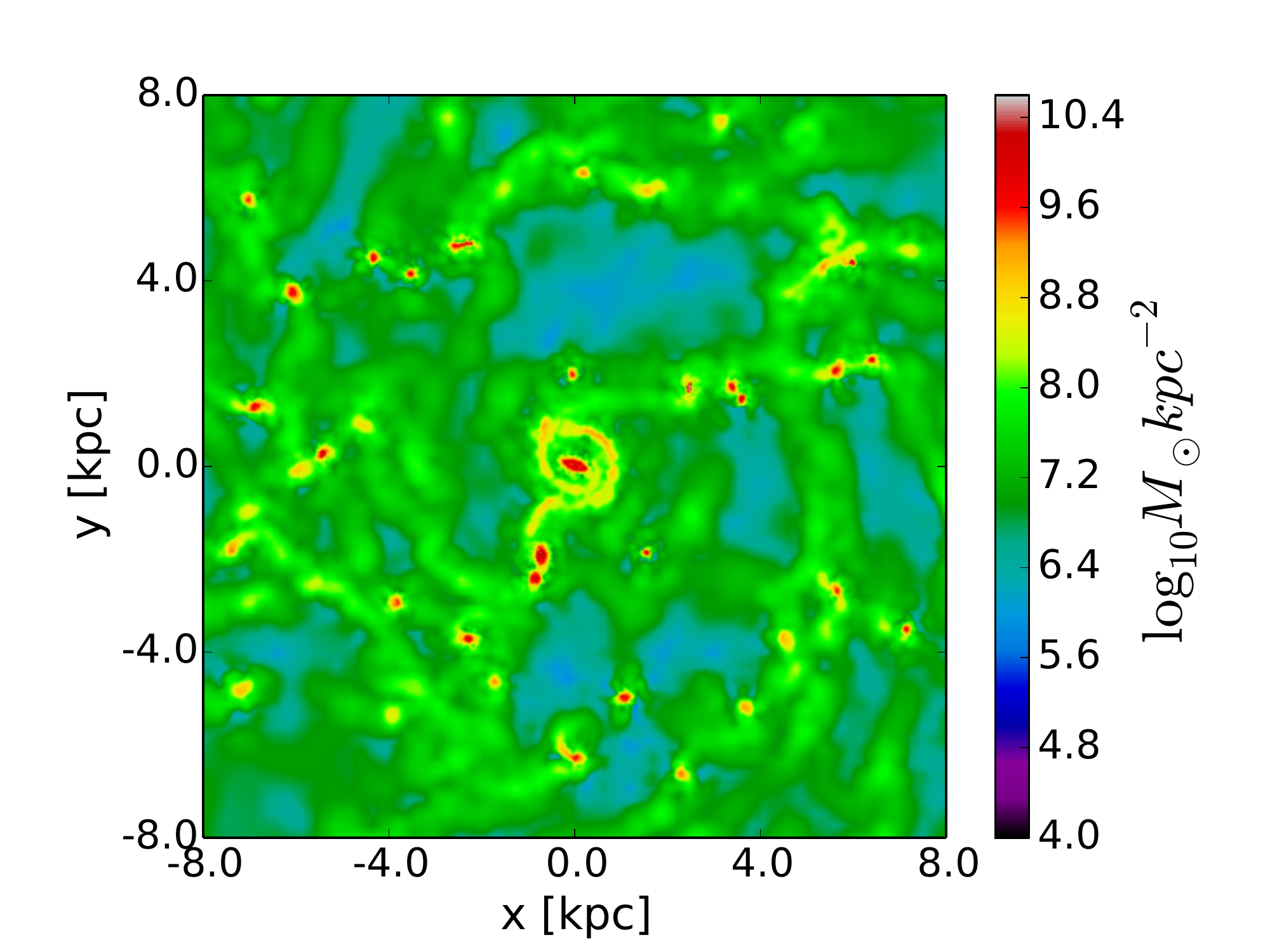}
  \includegraphics[trim={1cm 0 0 0}, width=0.33\textwidth]{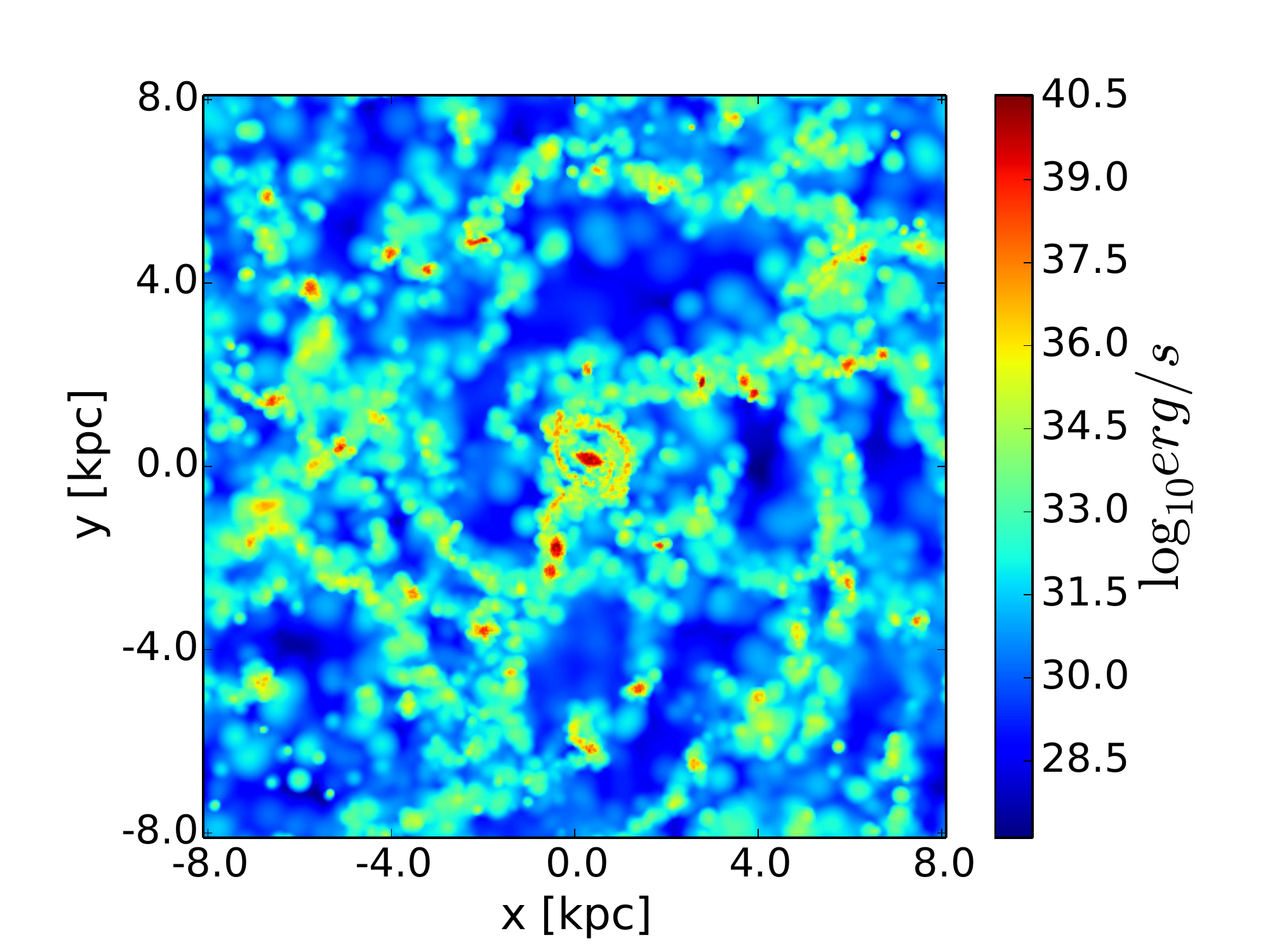}
  \includegraphics[trim={1cm 0 0 0}, width=0.33\textwidth]{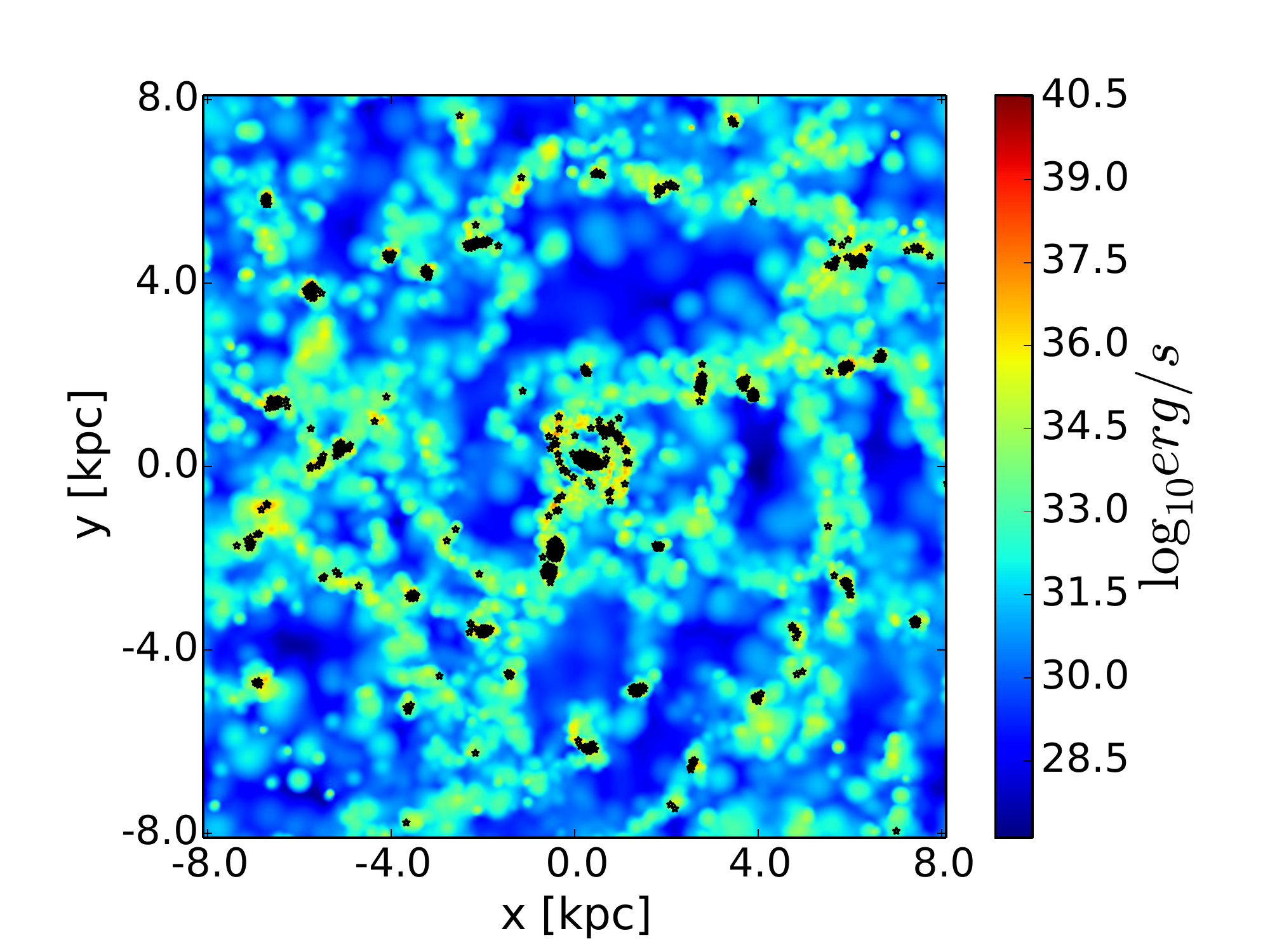}\\
  \caption{From left to right: gas density map, H$\alpha$ luminosity map obtained by postprocessing simulation with radiative transfer and H$\alpha$ luminosity map with young stars (age < 5~Myr, black stars). All the result are for a clumpy galaxy simulation at 200~Myr after the beginning of the simulation. H${\alpha}$ clumps with sizes $\sim 100$~pc are already present and closely trace star forming regions.}
  \label{fig:GasHalphaStars}
\end{figure*}
%
\begin{figure*}
  \includegraphics[width=0.49\textwidth]{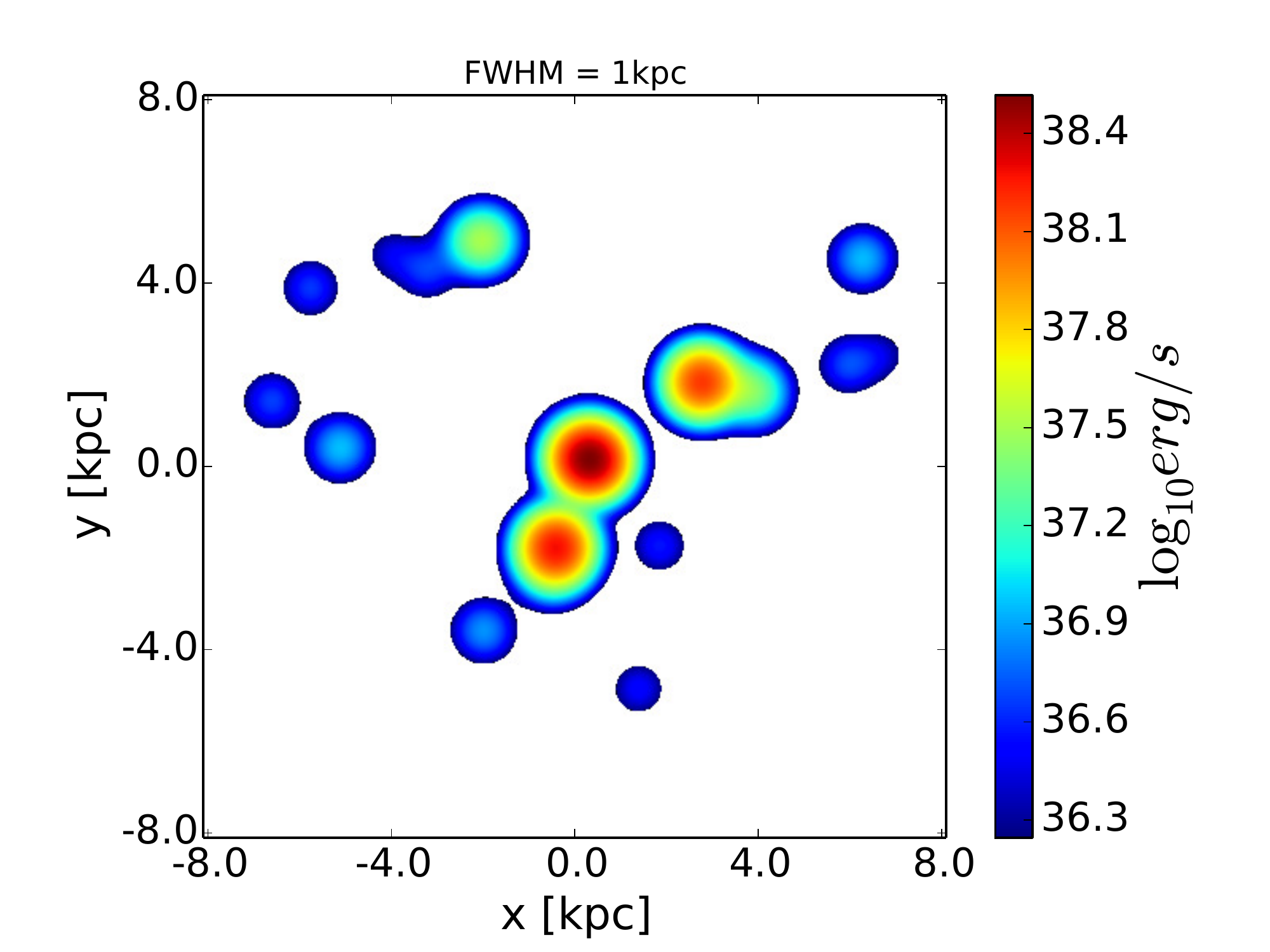}
  \includegraphics[width=0.49\textwidth]{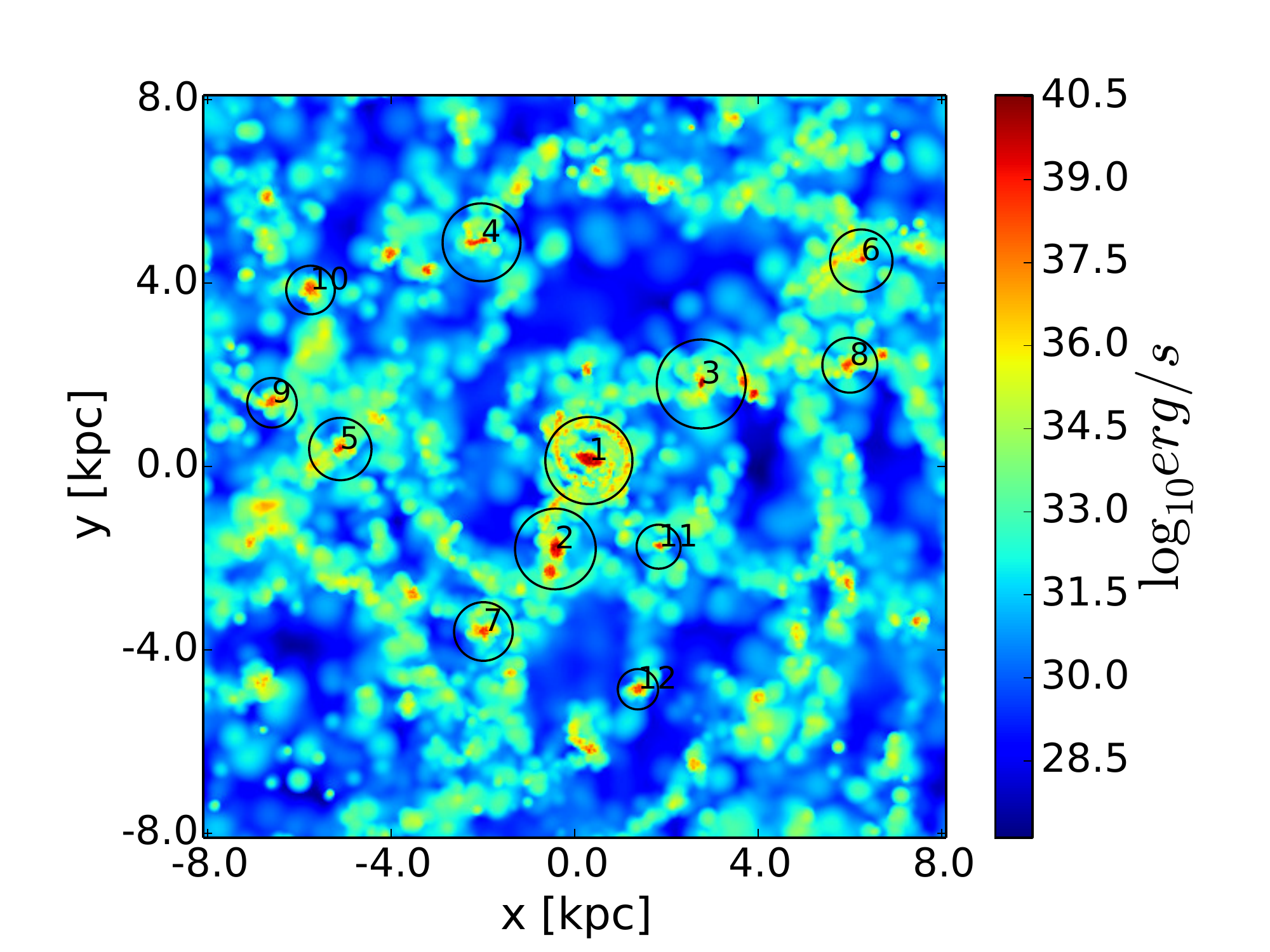}\\
  \includegraphics[width=0.49\textwidth]{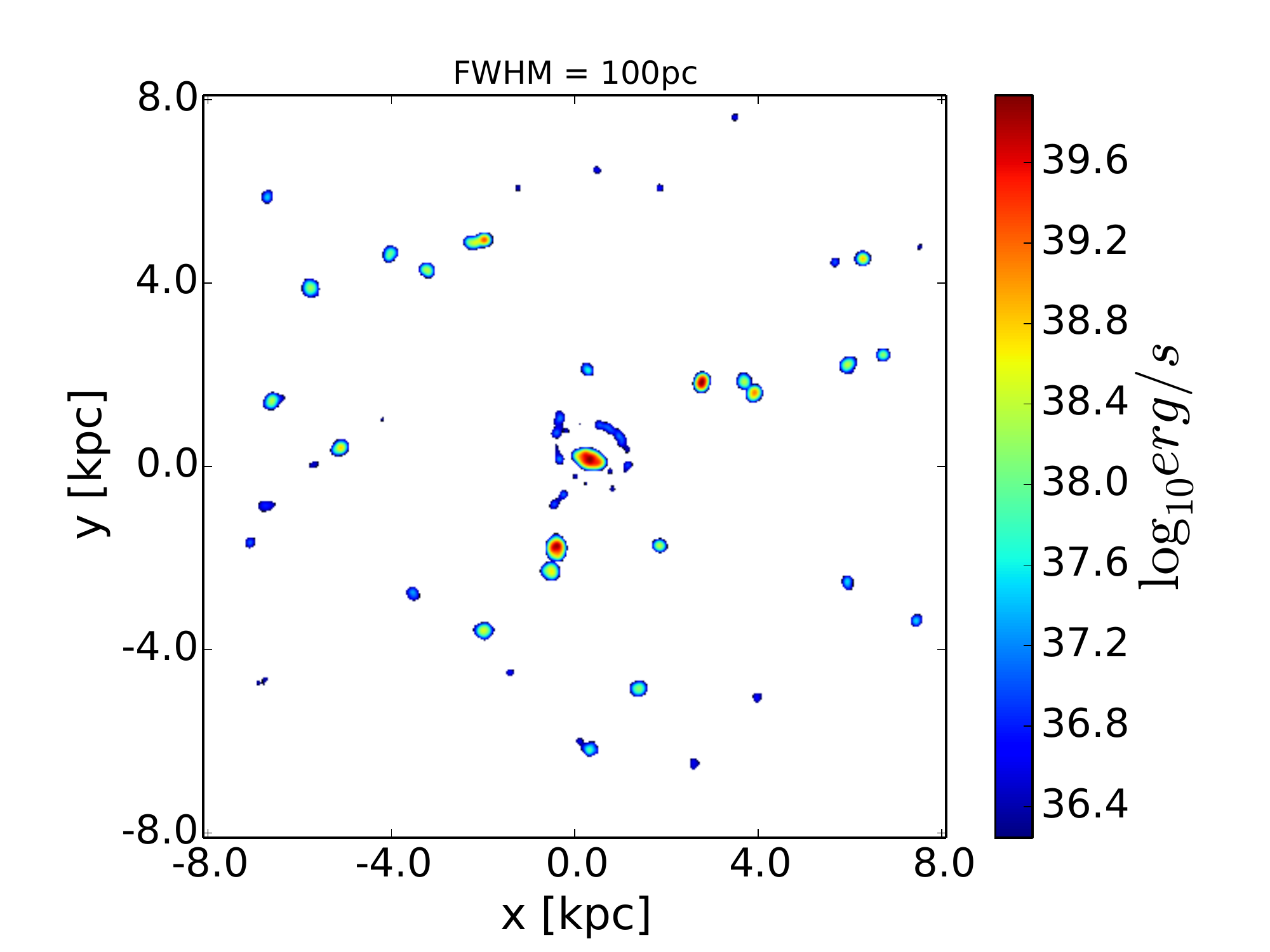}  
 \includegraphics[width=0.49\textwidth]{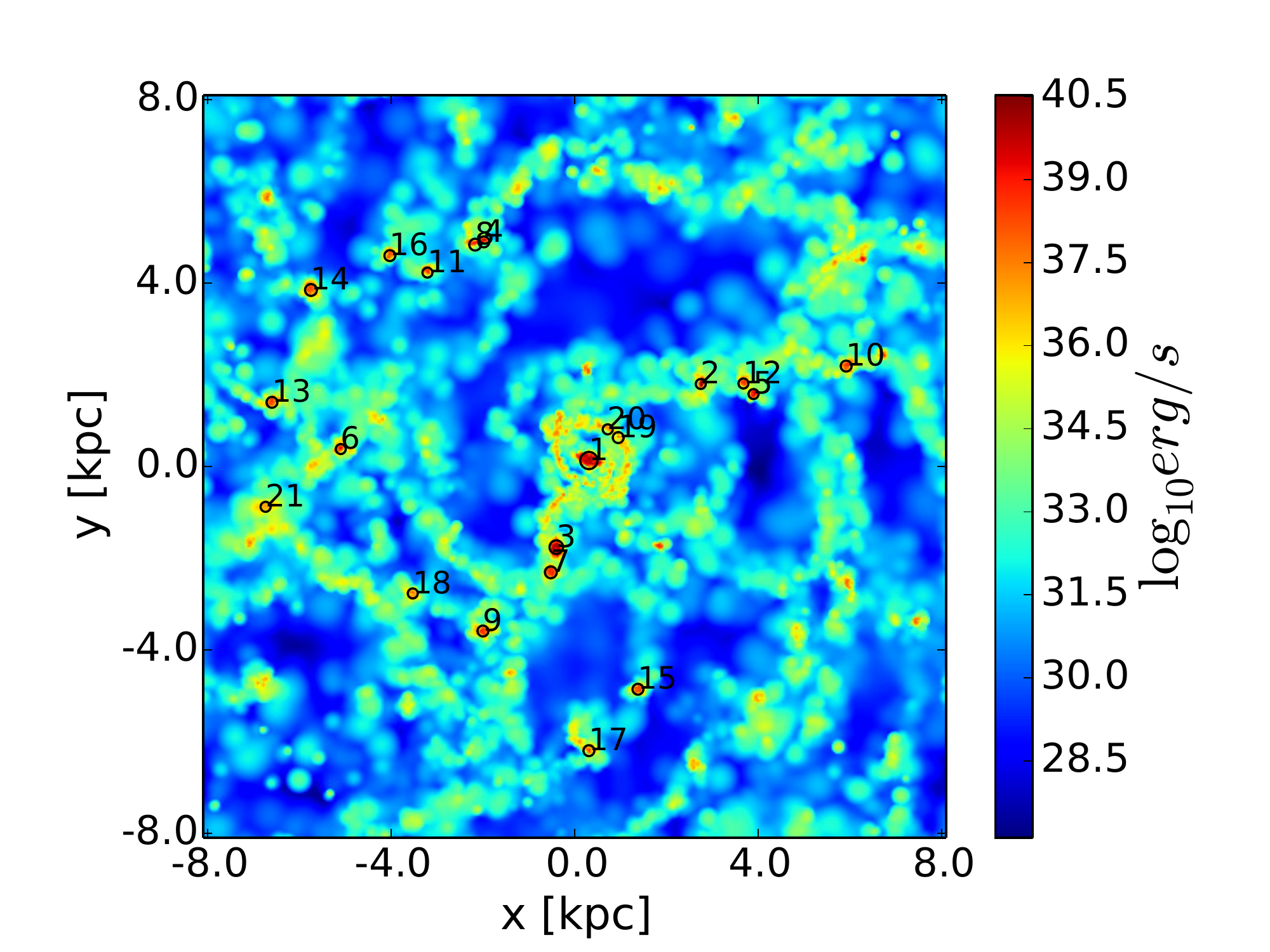}
  \caption{The left column shows identified clumps in the simulated $\alpha$ luminosity maps for the clumpy run at 200 Myr using $1$~kpc (top) and $100$~pc (bottom) smoothing length and an error level of $10 \%$ (after subtracting the background). On the right column the original simulated H$\alpha$ luminosity maps are shown with identified clumps in the convolved maps overplotted using circles. Lowering the spatial resolution makes the clumps appear larger.}
  \label{fig:comparison}
\end{figure*}
In this paper we focus on mocks of high-resolution hydrodynamical non-cosmological simulations of clumpy high-z gas-rich discs already presented in \citet{Tamburello2015}. 
The detailed structural properties  of all galaxy models are described in Table~2 of \citet{Tamburello2015}.
From the \citet{Tamburello2015} sample we chose the simulation that undergoes more fragmentation and gives rise to the largest clumps while including the effect of feedback. This simulation has the initial stellar mass of $4 \times 10^{10}$~M$_{\odot}$, embedded in a halo of $2.5 \times 10^{12}$~M$_{\odot}$, making the galaxy models
consistent with the stellar mass-halo mass relation predicted using abundance matching and other observational constraints \citep{Behroozi2013}.

The initial gas fraction is 0.5 and the spatial resolution (the gravitational softening length) is 100~pc. The simulation is performed with the N-body + smoothed particle hydrodynamic (SPH) code {\scshape gasoline}. In order to form a star, a gas particle has to have temperature lower than $3 \times 10^4$~K and density higher than $10$~cm$^{-3}$. The star formation efficiency parameter is $\epsilon_{SF} = 0.01$ and $4\times 10^{50}$~erg thermal energy is released into the ISM per supernovae event, shutting-off the cooling of the neighboring gas particles for a few tens of Myr (see \citealt{Stinson2006} and \citealt{Tamburello2015}).
This galaxy results in a highly clumpy structure (approximately after 100~Myr) and resembles $z \sim 2$ star forming galaxies. A key aspect for our purpose is that
this simulation produced clumps that are smaller and less massive compared 
to most previous work on massive disc fragmentation, 
having radii $\sim 100-300$~pc and masses in the range $10^7 - 10^8$~M$_{\odot}$, with some rare exceptions that can reach up to $10^{9}$~M$_{\odot}$ \citep{Tamburello2015}.

As will be explained in the next section, to produce H$\alpha$ maps, we post-process them by performing radiative transfer simulations to accurately calculate the  gas ionization state. 

\subsection{Radiative transfer and mock H$\alpha$ maps}

Our simulation is post-processed with the radiative transfer (RT) code {\scshape traphic} \citep{PawlikSchaye2008, Pawlik2011} 
to account for ionizing radiation from young stars and the meta-galactic ionizing radiation. For the background radiation we adopt a hydrogen photoionization rate of $10^{-12}$~s$^{-1}$ which is appropriate for redshift $z \sim 1-3$ \footnote{notice that also changing this number up to 10 times lower values, appropriate for higher and lower redshifts, does not change the results.}. The effect of UVB on ionization is accounted for by a density dependent self-shielding correction following \citealt{Rahmati2013}. 
For stellar radiation we assume all stellar particles younger than $5$~Myr are contributing to hydrogen ionizing radiation all with identical photon production rates of $6 \times 10^{46}$ photons per second per unit solar mass. In other words, we neglect the metallicity dependence and precise age dependence of stellar emissivities. The choice for stellar ionizing emissivities is motivated by results from {\scshape starburst99} \citep{Leitherer1999} and varies by a factor of a few around our choice depending on assumed metallicity and/or IMF \citep{Rahmati2013b}. 
For the energy distribution of ionising radiation, we assume a black-body radiation with an effective temperature of $4 \times 10^4 K$. 
Then, sources can emit ionizing radiation and the RT code {\scshape traphic} follows the propagation of ionizing radiation in photon packets which travel with the speed of light, and the ionization state of gas in post-processing (i.e., the distribution of gas and sources are fixed during the RT simulation). Then the RT simulation proceeds until the ionization equilibrium is reached. Collisional ionization is also accounted for.

After calculating the ionization state of the gas, we used SPH interpolation to make 2D H$\alpha$ maps of the simulated galaxy. In order to have the same resolution we have in \citet{Tamburello2015} ($\sim 10$~pc SPH smoothing length), we fixed the cell size at $10$~pc.
We calculate the intensity of H$\alpha$ emission for each pixel using:
\begin{equation}
I_{\nu} (H\alpha) = \frac{h\nu(H\alpha)}{4\pi} \int n_e~n_p~\alpha(H{\alpha})~dx,
\end{equation}
where the integration is performed along the line-of-sight (projection direction). For the temperature dependent H$\alpha$ emission coefficient we use $\alpha(H{\alpha}) = 7.86 \times 10^{-14} \times 10^4/T~cm^3s^{-1}$, following \citet{Osterbrock2006}. We note that normalising the total luminosity of our mock maps with the star formation rate of the galaxy, using the following relation from \citealt{Kennicutt98}
\begin{equation}
\rm{SFR (M_{\odot}~yr^{-1})} = \frac{L(H\alpha)}{1.26 \times 10^{41} ergs~s^{-1}}
\label{eq:kennicutt}
\end{equation}
produces the same results. We do not include dust estinction, whose precise modelling is beyond the scope of this work. 
\subsection{Identifying the clumps}\label{s:findClumps}
After producing the H${\alpha}$ maps, in order to mimick observations, we add errors equivalent to $1 \%$, $5 \%$ and $10 \%$ of the total H${\alpha}$ luminosity, 
uniformly distributed among all pixels. We accept as signal everything that is above $5 \times n$, where n is the error per pixel, in order to be consistent with observations (sometimes a threshold of $3 \times n$ is used, but we check that our results do not change). The background (every signal lower than $5 \times n$) is set to zero for simplicity. Finally, in order to mimick the instrumental resolution effects, we convolve the H$\alpha$ maps with a 2D Gaussian aperture with different values of FWHM: 100~pc, appropriate for comparison with clumps found in lensed galaxies, and 1~kpc, which is comparable to the typical resolution accessible for HST observations at $z \geq 1$. 
To identify the clumps in our H$\alpha$ luminosity maps, we procede iteratively as follows. First of all, we find the maximum point of luminosity, $P_{max} = (x_{max},y_{max})$, in the map, then, by means of circles of increasing radius, $R_i$, around $P_{max}$, we analyse the integrated luminosity associated with the clump candidate, that is
\begin{equation}
L_i(R_i) = \int^{R_i}_0 l(r) dr 
\end{equation}
We set the clump radius $R_{clump}$ when either an area of background (set to zero) is reached, i.e. $l(R_{clump}) = 0$, or when the average radial luminosity density, defined as
\begin{equation} 
\int^{R_{clump}}_{R_{i-1}} l(r) dr /  (R_{clump}^2 - R_{i-1}^2)
\end{equation} 
is increasing, most likely related to the presence of another clump. Once a clump is found, it is removed from the map and we return to the first step. To prevent the possibility of having spurious clumps, we consider in our analysis only clumps which have radii ten times the cell size, so that the minimum clump radius is 100~pc. Moreover, for our analysis, we only consider clumps which have sizes equal or larger that the Gaussian aperture used for smoothing the maps (e.g. radius of $500$~pc for the smoothing using $FWHM = 1$~kpc).
%
%
%
\section{Results}
%
\begin{figure*}
  \includegraphics[width=0.49\textwidth]{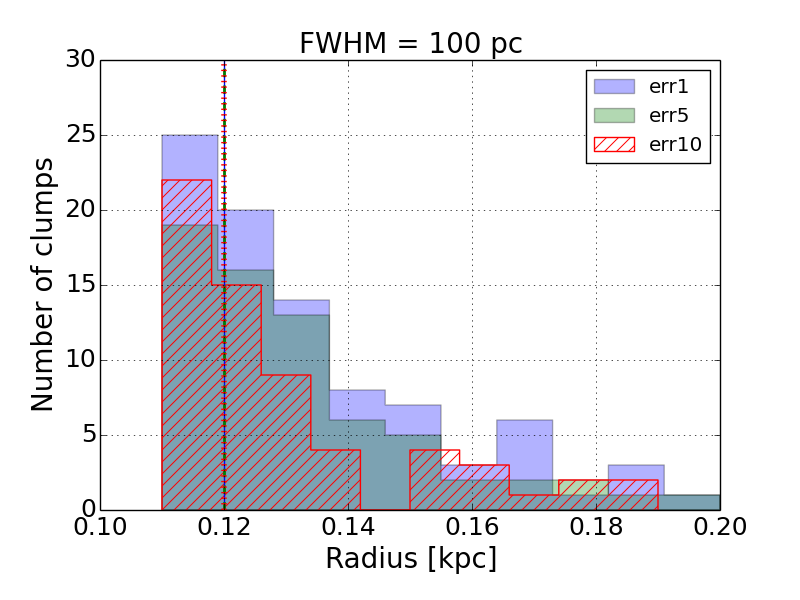}
  \includegraphics[width=0.49\textwidth]{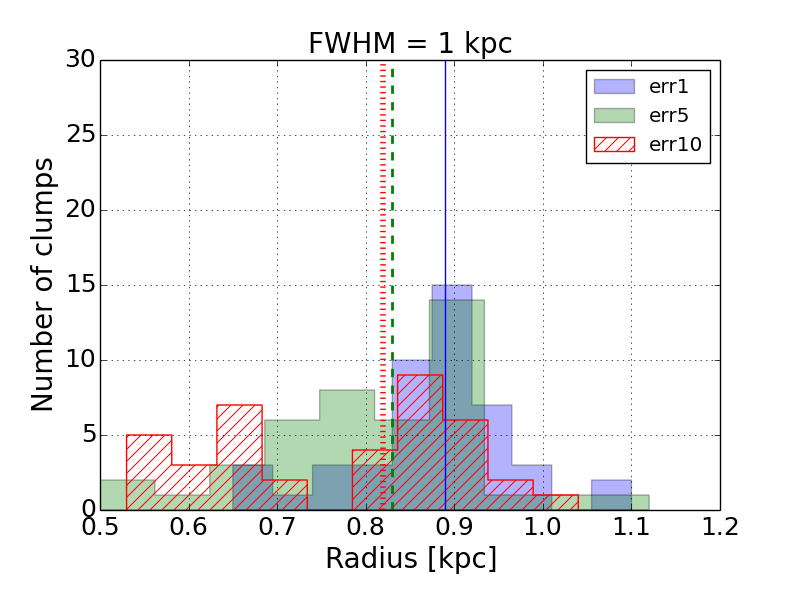}\\
  \includegraphics[width=0.49\textwidth]{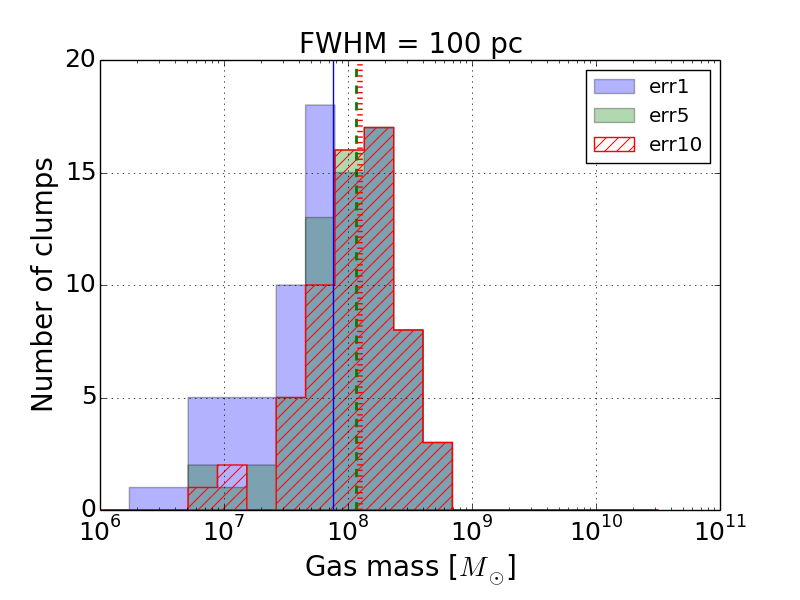}
  \includegraphics[width=0.49\textwidth]{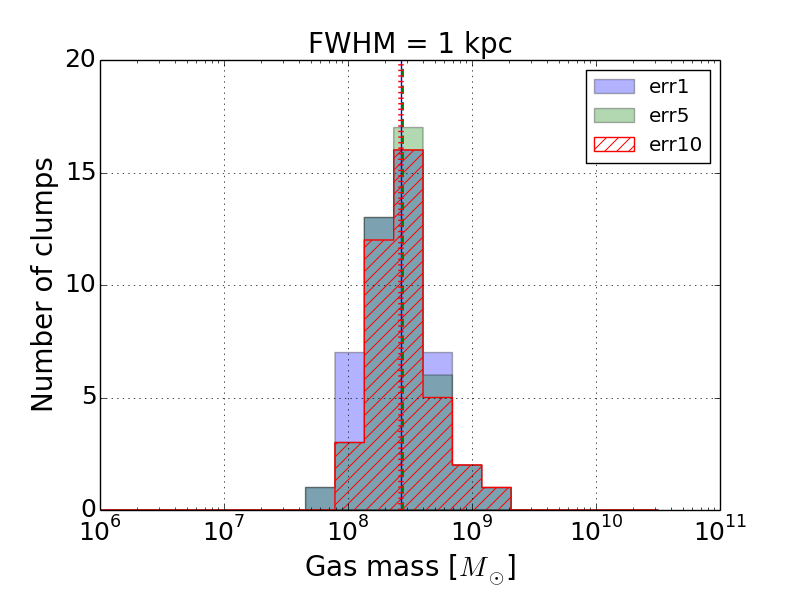}\\
  \includegraphics[width=0.49\textwidth]{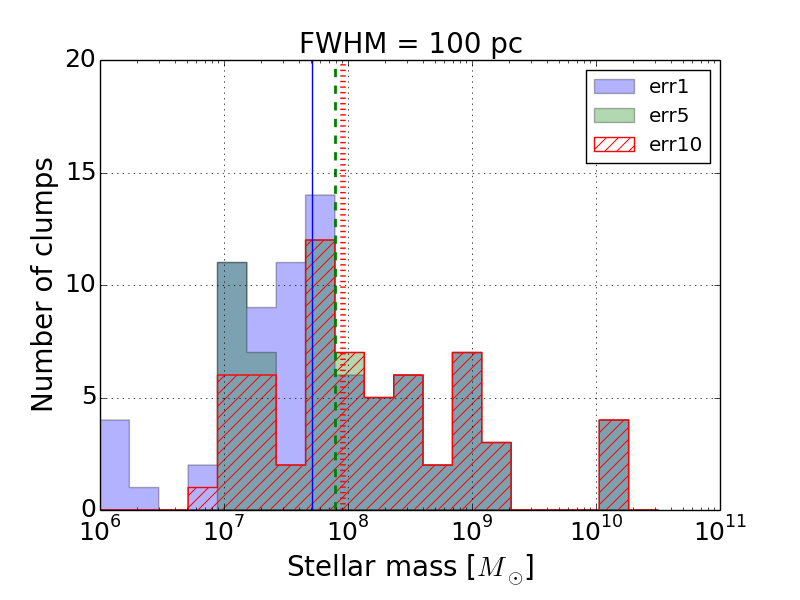}  
  \includegraphics[width=0.49\textwidth]{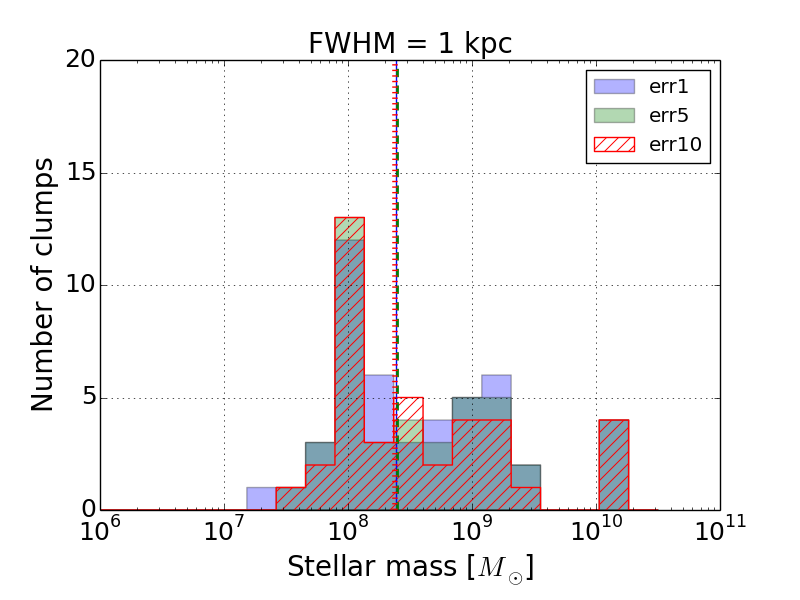}  
  \caption{From top to bottom we show the H$\alpha$ clump radius, total gas (neutral and ionized) mass and stellar mass distributions for our simulation combining different times (200, 300, 400 and 500 Myr) and using different level of error (i.e. 1$\%$, 5$\%$, 10$\%$ of the total luminosity). Distributions are obtained based on H$\alpha$ maps convolved with FWHM $= 100$~pc (left) and FWHM $= 1$~kpc  (right), using different levels of error. The medians of the distributions for the error levels of 1$\%$, 5$\%$ and 10$\%$ are shown using solid, dashed and dotted vertical lines, respectively. The results are weakly sensitive to the level of error. The clump sizes strongly vary  by changing the observed spatial resolution, while clumps masses (both stellar and gaseous) are less sensitive to it.}
  \label{fig:MassRadiusPlots}
\end{figure*}
%
\begin{figure*}
  \includegraphics[width=0.49\textwidth]{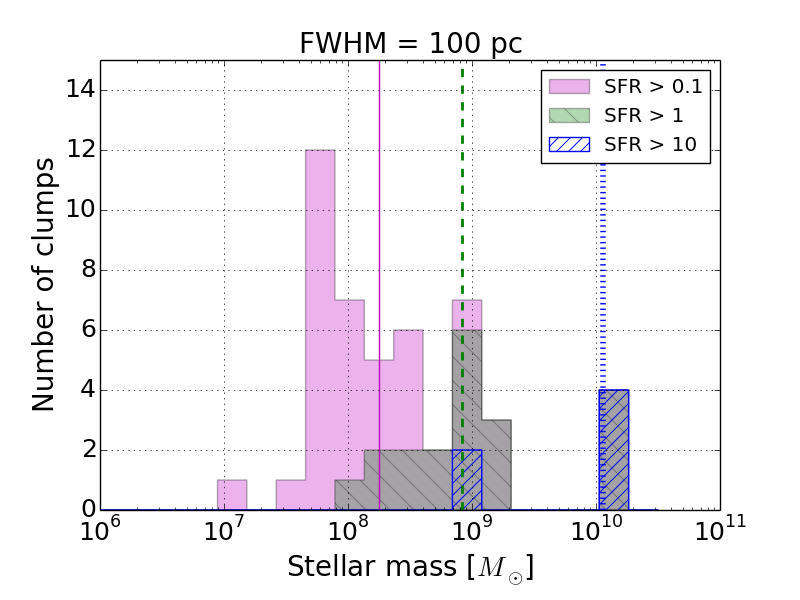}
  \includegraphics[width=0.49\textwidth]{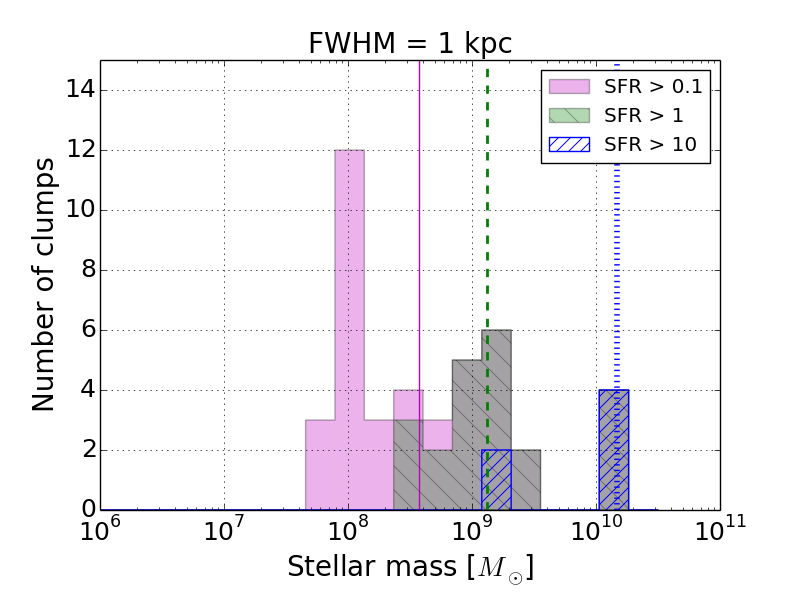}\\
  \caption{The sensitivity of stellar mass distribution of H$\alpha$ clumps to the detection threshold used for observing them. The mass distributions are shown for H$\alpha$  maps with 5$\%$ error levels and for two different spatial resolution (100 pc on the left and 1 kpc on the right) and after combining different times (200, 300, 400 and 500 Myr). For each resolution the mass distributions are shown for SFR thresholds of 0.1, 1 and 10 M$_{\odot}$/yr and the vertical solid, dashed and dotted lines show their corresponding medians. The detection threshold (sensitivity) strongly affects the mass distribution of the clumps irrespective of the spatial resolution.}
  \label{fig:SFRcuts}
\end{figure*}
Here we present our main results in order to investigate the effect that resolution has in determining the observed clump sizes and masses, and how these compare with the intrinsic clump sizes and masses in the simulations. 
The left diagram in Figure \ref{fig:GasHalphaStars} shows the simulated gas density map at 200~Myr. The middle panel shows H$\alpha$ map, while the right panel shows the same with the position of young stars (the main ionizing sources in our calculations) overplotted using black dots. Comparing these two images, one can see that gas clumps are star-forming regions, well tracked by H${\alpha}$. Therefore, while  H${\alpha}$ is a tracer of ionized gas, it also performs well as a tracer of star forming regions (young stars). 
In Figure \ref{fig:comparison} we show how the H${\alpha}$ luminosity map appears once we convolve it with a 2D Gaussian aperture with FWHM$ = 1$~kpc (top left panel) and FWHM$ = 100$~pc (bottom left panel). To show the sensitivity of H$\alpha$ clump properties to the underlying spatial resolution, in each map in Figure \ref{fig:comparison}, we also show the identified clumps with black circles.  
When clumps are selected in the convolved map with FWHM $ = 1 $~kpc, indeed, they appear larger, either because they merge together or flux is smeared out. However, identified clumps in the map with FWHM $= 100$~pc are well resolved, nearly as good as they are in the original simulated maps without any smoothing. Moreover, the number of clumps in higher resolution map is nearly twice the number of clumps identified in the lower resolution map.
%
\begin{figure}
  \includegraphics[width=0.49\textwidth]{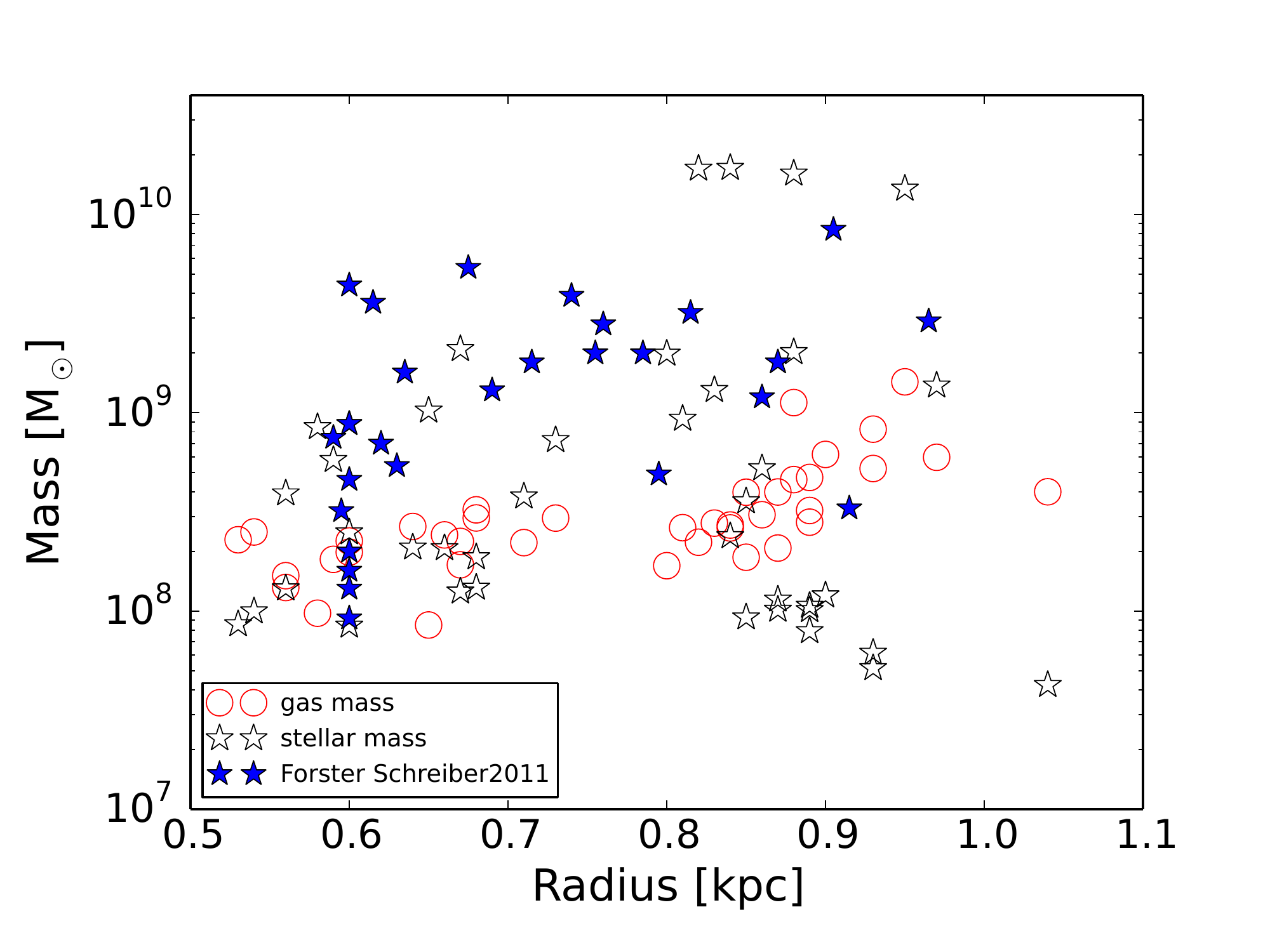}\\
  \caption{The simulated clump gas/stellar mass versus radius relation (shown with circle/open star symbols) compared with the observed clump stellar mass vs. radius distribution from \citet{FoersterSchreiber2011b} (shown using filled star symbols). The simulated clumps are identified in mock H$\alpha$ maps smoothed with a Gaussian aperture of $1$~kpc, comparable to the typical resolution of the observed sample (with PSF $FWHM = 1.2$~kpc in the HST NICMOS/NIC2 H$_{160}$ band). The gas and stellar clump masses are comparable and in good agreement with observations at similar radii.}
  \label{fig:MassVsRadius}
\end{figure}
We repeat this analysis at different times (200 - 300 - 400 - 500~Myr), using different error values and convolution lengths. At each time we compute the H$\alpha$ clump radii and masses. 
Combining all the time steps, the distribution of H$\alpha$ clump radii is shown in the top row of Figure \ref{fig:MassRadiusPlots}. We show the results for both the resolutions obtained using FWHM $= 100$~pc (on the left) and FWHM~$= 1$~kpc (on the right). Clumps selected after a Gaussian convolution with FWHM~$= 100$~pc have sizes $\sim 100$~pc as we previously showed in \citet{Tamburello2015}, but their sizes increase up to 1~kpc when we make use of the Gaussian convolution with FWHM $= 1$~kpc. The result is not too sensitive to the level of error, however the larger the error is, the fewer clumps we are able to find.

The clump gas mass distributions using FWHM $= 100$~pc (on the left) and FWHM~$= 1$~kpc (on the right) and different errors are shown in the second row of Figure  \ref{fig:MassRadiusPlots}. Here we define the clump mass as the total (neutral and ionized) gas within the clump radius, using a 2D gas density map. We also measure the stellar mass for each clump in the same way, by measuring the total stellar mass within an aperture defined by the clump radius in H$\alpha$ maps. This is very similar to the method used by \citealt{FoersterSchreiber2011b} to calculate the stellar masses of clumps identified in their H$_{160}$ band observations.
We find fewer, but more massive (by a factor of $\sim 2$) clumps with FWHM~$= 1$~kpc, compared to the case with FWHM~$= 100$~pc both for gas and stellar mass estimates. The peak of both gas and stellar mass distributions for the lower resolution case is at $\sim 2 \times 10^8$~M${_{\odot}}$. The sensitivity of the mass distributions to the spatial resolution is not as strong as that for the clump radius distribution. While the typical clump radius becomes 10 times larger by changing the spatial resolution from FWHM~$= 100$~pc to FWHM~$= 1$~kpc, the typical clump mass changes only by a factor of 2. This is due to the fact that the convolution simply spreads out the same amount of mass/luminosity over a much larger area whick strongly 
affects their measurable sizes. However, since most of the mass in the larger area is still inside the intrinsic radius of the clumps, the effect on the measured masses is not expected to be large. Nonetheless, when two massive clumps are relatively close, as it is often the case since clump-clump mergers are common, the larger area would encompass the sum of their mass, which explains the factor of $\sim 2$ increase in the mass distribution of the clumps as we decrease the spatial resolution. 

The observed masses of the clumps could be more affected by another feature of the observations, namely their sensitivity limit. At any given sensitivity only objects brighter than the corresponding detection threshold are observed. Figure \ref{fig:SFRcuts} shows how using three different detection thresholds (SFR~$>$~0.1, 1, 10 M$_{\odot}$/yr) results in drastically different clump stellar mass distributions irrespective of the used spatial resolution. Changing the sensitivity within the aforementioned range by a factor of 10 translates into nearly one order of magnitude change in the typical clump stellar mass. For a fixed detection threshold, however, changing the resolution from 100~pc to 1~kpc changes the typical clump stellar mass by a factor of $\lesssim 2$. This is in agreement with what found in (Dessauges et al. in prep.) where the sensitivity seems to have a stronger impact on stellar mass distribution than the spatial resolution.

It is important to recall that, unlike simulations, observations do not have direct access to the actual mass inside the clumps and indirect
methods should be used to estimate the observed clump masses. Such methods often are based on specific theoretical assumptions about the nature of the clumps, such as assuming that clumps are bound and/or virialized objects. In these cases the measured masses would be significantly inflated simply because they are derived using equations that express the mass as a function the observed radius, which, as we have shown, can be overestimated by an order of magnitude. For illustration, we can adopt the notion that clumps are virialized. In this case the clump mass, $M$, radius, $R$, and velocity, $v$, are connected by the relation $M = v^2R/G$, where $G$ is the gravitational constant. The velocity estimations are often based on high resolution spectra and they are less prone to uncertainties related to the limited spatial resolution, so to
be conservative we will assume they can be estimated correctly. Our clumps, having radii of $\sim 100$~pc, could be seen as kpc-size objects (see the first row in Figure \ref{fig:MassRadiusPlots}), which leads to their masses being over-estimated by one order of magnitude. As a result, a typical clump in the mass range of $\approx 10^8$~M$_{\odot}$ and radius range of $\sim 120$~pc, would be seen as a kpc-size clump with an overestimated mass of $\sim 2 \times 10^9$~M$_{\odot}$, while directly measuring the mass inside the new aperture (size of the clump using low resolution) results in $\approx 3 \times 10^8$~M$_{\odot}$. Some published works have also assumed that the observed clump radius is proportional to the Jeans length $\lambda_J$ \citep{Wisnioski2012}, which would imply that its mass is of order the Jeans mass $M_J=
{(\lambda_J/2)}^3 \times \rho_g$ where $\rho_g$ is the mean local gas density in the disc. This results in the clump mass to scale as $\lambda_J^3/{2}^3 \sim (\lambda_J)^3/8$. With the latter scaling, even if the radius is overestimated by only a factor of 2, the correspondin mass
will be over-estimated by about an order of magnitude.

The clump mass versus radius distribution is shown in Figure \ref{fig:MassVsRadius}. Stars mark the clump stellar mass estimation from observations (filled stars; \citealt{FoersterSchreiber2011b}) and from our simulation (empty stars). Those observations are obtained with the resolution equivalent to $1.2$~kpc (the FWHM of the point spread function in HST NICMOS/NIC2 H$_{160}$ band). To match this as closely as possible, we use our clump mass/size estimations after using a smoothing length of $FWHM = 1$~kpc. Circles show the total (neutral and ionized) gas mass distribution of the clumps estimated from our simulation (empty circles). Both the clump stellar and gas mass vs. radius distributions are in good agreement with each other and with the observed clump stellar mass vs. radius distribution. Note that the clump stellar masses and sizes calculated by identifying gravitationally bound structures (using {\scshape SKID} group finder, \citealt{Stadel2001}) and without convolving them with a proper Gaussian filter (presented in \citealt{Tamburello2015}) would fall in the left bottom corner of Figure \ref{fig:MassVsRadius}, with clump radii and masses spanning ranges from $\sim 100-300$~pc and $10^7$ to $10^9$~M$_{\odot}$, respectively. These radii and masses would not be consistent with observations at  face value, but after convolving them using the proper spatial resolution, they match the observed data well.

We also note that the clumps observed by \citet{FoersterSchreiber2011b} are detected in stellar continuum maps, while we use H$\alpha$ mock observations to identify our clumps. However the similarity between the clump mass/size distributions we found in our earlier analysis, where we identified gravitationally bound stellar structures as clumps, and those we found here by identifying H$\alpha$ clumps, justifies the comparison illustrated in Figure \ref{fig:MassVsRadius}. We postpone a more precise comparison to future work in which we will present multi-band mocks of the stellar continuum.

\begin{figure}
  \includegraphics[width=0.49\textwidth]{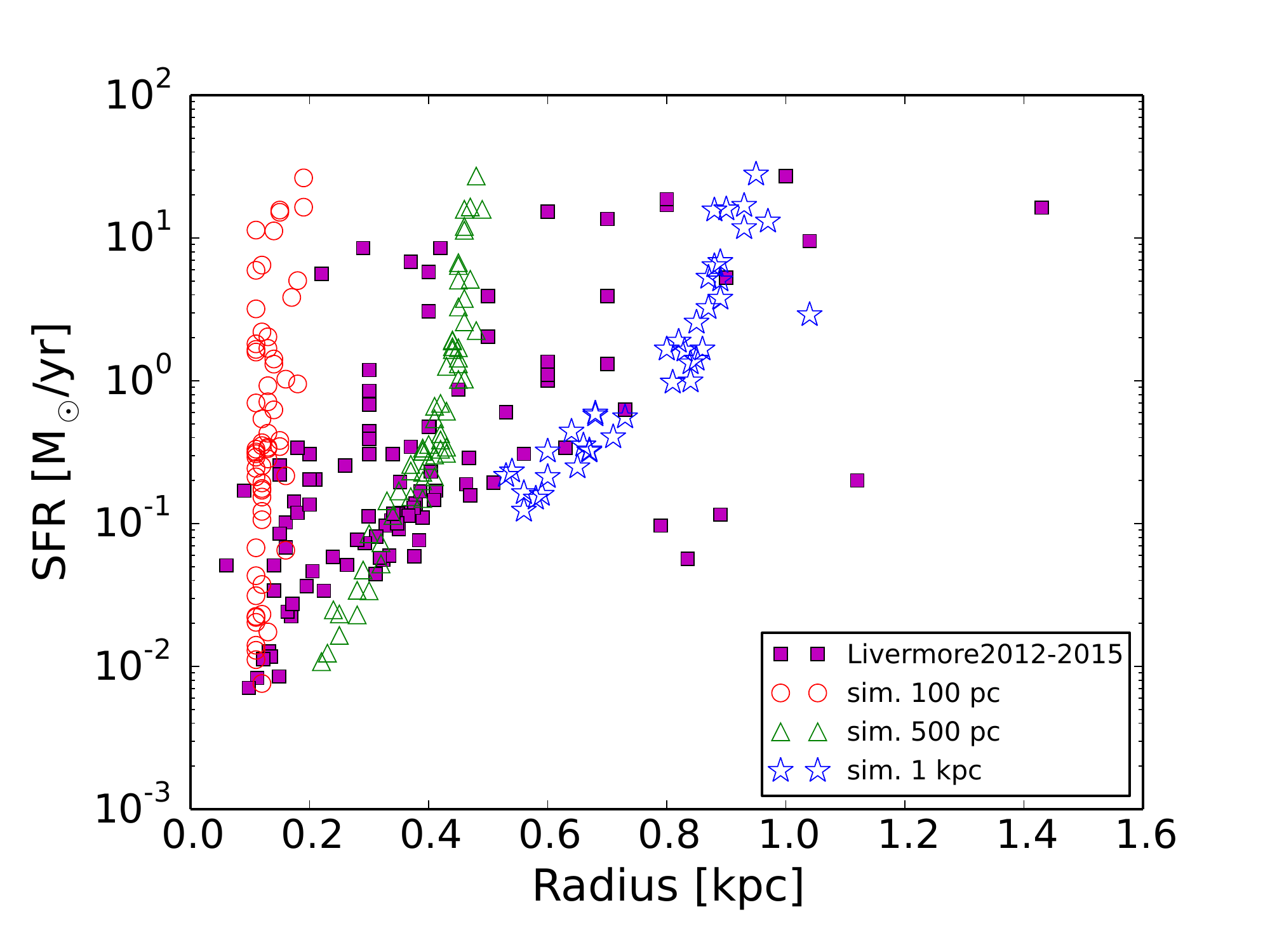}\\
  \caption{The comparison between clump SFR vs. radius distributions in simulations (empty symbols) and observations (filled squares). Observational data points are for lensed galaxies from \citet{Livermore2012, Livermore2015} where the spatial resolution varies between 100 pc and 700 pc. The simulation results are shown for three different spatial resolution: 100 pc, 500 pc and 1~kpc, shown with red circles, green triangles and blue stars, respectively. The combination of simulations with different resolution and observations which have a similar range of spatial resolution match very well. This suggests a correlation between clump radius and the spatial resolution used for observations.}
\label{fig:SFRvsR}
\end{figure}

In Figure \ref{fig:SFRvsR} we compare simulated clump SFR vs. radius with observational data of lensed galaxies from \citet{Livermore2012, Livermore2015}, after multiplying their reported SFRs by 1.7 required to make them consistent with the Salpeter initial mass function \citep{Salpeter55} assumed in equation \ref{eq:kennicutt}. Noting that the spatial resolution used for the observations (magenta squares) varies between 100 pc 
and 700 pc, we compare them with simulation results using three different spatial resolution: 100 pc, 500 pc and 1 kpc (red circles, green triangles and blue stars, respectively). The combination of simulations with different resolution and observations which have a similar range of spatial resolution match very well. This suggests a correlation between clump radius and the spatial resolution used for observations.

\begin{figure}
  \includegraphics[width=0.49\textwidth]{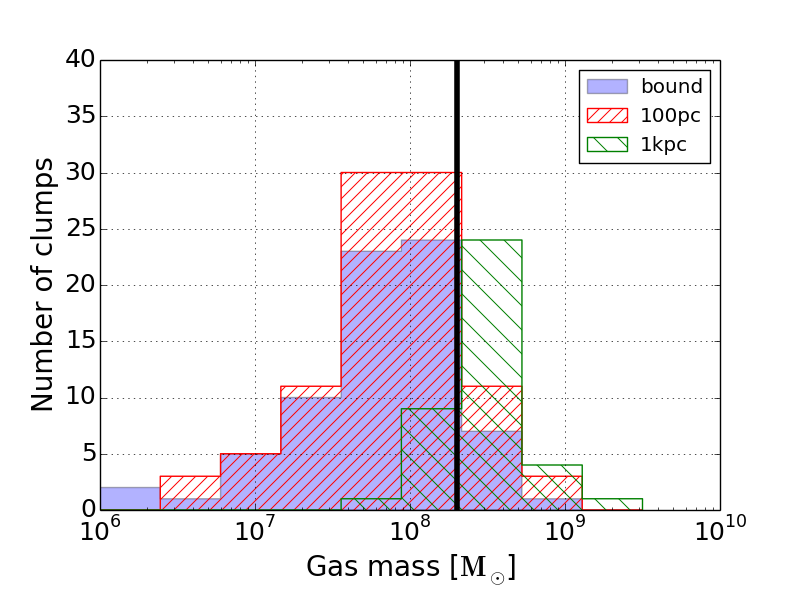}\\
  \caption{Comparison between the clump gas mass distribution using gravitationally bound structures in the simulation (filled histogram) and those found for the clumps in the H$\alpha$ intensity maps using two different resolutions of 100 pc and 1 kpc (hatched histograms). The results are obtained by combining the simulation outputs at different times (i.e. 200, 300, 400 and 500~Myr). The black vertical line shows the Toomre mass estimated at the onset of the disc  fragmentation, which coincidentally agrees well with the inflated clump mass distribution we found with the lower 1 kpc resolution.}
  \label{fig:MassSelVsBound}
\end{figure}

Since in this paper we base our analysis only on $H{\alpha}$ mocks, it is reasonable to ask how well the identification of clumps via $H{\alpha}$ can capture the intrinsic properties of clumps, namely their radii and masses, when high resolution is available. This is important to address in order to conclude that the trends that we have presented here are meaningful. A direct comparison between the clump gas mass estimated using bound structures in the simulation (using {\scshape SKID} group finder; see \citealt{Tamburello2015}) and those defined by identifying the clumps in the 
H$\alpha$ intensity maps is shown in Figure \ref{fig:MassSelVsBound}, where we combine the simulation results at different times (i.e. 200, 300, 400 and 500~Myr). The mass distribution of the gravitationally bound clumps (filled histogram) is very similar to the mass distribution of H$\alpha$ clumps smoothed with 100~pc resolution (red hatched histogram). Using a lower resolution of 1~kpc for smoothing the H$\alpha$ maps results in a different distribution which peaks at a higher mass (green hatched histogram). Finally, the solid vertical black line in Figure \ref{fig:MassSelVsBound} indicates the Toomre mass calculated for our simulation at the onset of disc fragmentation, using the local gas disc properties (see Tamburello et al. 2015). Interestingly, this agrees well with the inflated clump mass distribution we found with the lower 1~kpc resolution. However this minimum Toomre mass estimation only has small overlap with the high mass tail of the clump mass distribution using the higher 100~pc resolution. This highlights how using common notions and definitions of the fragmentation scale, taken from simple linear perturbation theory, can be extremely misleading, producing an apparent match with the data because of lack of resolution. This is not surprising as nonlinear phenomena, such as disc fragmentation, are hardly described by linear theory, as discussed in detail elsewhere \citep{Boley2010, Tamburello2015}. The characteristic fragmentation mass in the simulation, which is recovered at higher spatial resolution, is much better matched with the modified Toomre mass proposed by \citet{Tamburello2015}, which takes into account some of the nonlinear aspects of the process.
%

%


\section{Summary and discussion}
Studying the nature of clumps observed in high-z galaxies has been subject to active theoretical work. However, there are some sensitive aspects which complicates the direct comparison between observations and model predictions. For instance, most theoretical studies use gas density distribution to identify clumps or interpret their properties, while observational studies are primarily based on detecting ionized gas and/or stellar emission, which does not necessarily trace the gas density. Moreover, the finite spatial resolution and sensitivity of observational studies should be taken into account when comparing with simulations which often have different resolutions. In this work, we tackled these two problems by post-processing hydrodynamical simulations of clumpy galaxies with accurate radiative transfer calculation which allows us to identify clumps in H$\alpha$ maps, similar to what is done in observational studies. In addition, we convolved the simulation outputs with different spatial resolutions in order to investigate the impact of resolution on the resulting clump properties. 

We found that density peaks and peaks in H$\alpha$ emission do not always coincide with each other, mainly because H$\alpha$ emission is sensitive to both the gas density distribution and the distribution of young stars which are the main ionizing sources. The differences, however, are not large and the mass distribution of clumps found from H$\alpha$ maps are very similar to what one finds by identifying gravitationally bound objects considering only the gas density distribution \citep[e.g.,][]{Tamburello2015}. This shows that, provided that enough resolution and sensitivity is available, H$\alpha$ observations are a very powerful tool to study star
forming clumps and extract their physical properties in a meaningful way.

We found that the spatial resolution have a large impact on both mass and size of the clumps. Smoothing our simulate H$\alpha$ emission maps with Gaussians with FWHM of 100 pc and 1 kpc resulted in median clump masses $1.1 \times 10^8$ and $2.6\times 10^8~\rm{M_{\odot}}$, respectively. We found that the clump sizes are more sensitive to the spatial resolution where using smoothing lengths of 100 pc and 1 kpc resulted in median clump radii of 120 pc and 900 pc, respectively. Moreover, we found that varying the level of the error we added to our mock H$\alpha$ maps, namely 1, 5 and 10$\%$ of the total luminosity, does not affect the distribution of clump sizes and/or masses.

In addition we tested how the sensitivity (detection threshold) used for observing the clumps changes their observed properties. Changing the sensitivity by a factor of 10 (e.g., from SFR $>$ 1 to SFR $>$ 10~M$_{\odot}$/yr) increases the typical clump stellar mass by nearly one order of magnitude. For a fixed detection threshold, however, changing the resolution from 100~pc to 1~kpc increases the typical clump stellar mass only by a factor of $\lesssim 2$. This result is also supported by a careful analysis of the measured clump stellar masses where data with different sensitivities and spatial resolutions are combined (Dessauges et al. in prep.).

The sensitivity of the clump properties to the spatial resolution and the detection threshold used to observe them has profound consequences for comparing models and observed data. For instance, as we showed in \citet{Tamburello2015}, selecting clumps using the gas and stellar density distributions results in typical clump sizes in the range $\sim100$ pc, while high-z observations indicate typical clump sizes more than 10 times larger \citep[e.g.,][]{FoersterSchreiber2011b}. At the first glance this might indicate inconsistencies between model predictions and observations, but as we showed here, after convolving our simulated H$\alpha$ maps with a smoothing length similar to those typically accessible to high-z observations (i.e., $\gtrsim 1$ kpc), we found clump sizes and masses very similar to those claimed by observational studies. This conclusion is in line with recent observations that used gravitational lenses to achieve a spatial resolution of $\sim 100$ pc and found clumps much smaller than those typically found in high-z galaxies observed with $\sim$ kpc resolution \citep{Jones2010, Livermore2012, Livermore2015}. These high-resolution observations found clump sizes very similar to what we found for our simulated H$\alpha$ maps using a smoothing length of $100$~pc. This conclusion is also in good agreement with what \citet{Fisher2016} recently found for lower redshift disc galaxies with prominent substructures.

Finally, it is important to note that the incorrect derivation of clump physical properties because of different observational limitations can lead to incorrect inferences on their nature, such as verifying that the observationally implied clump sizes match a fragmentation scale predicted by simple theoretical models that miss nonlinear effects. This apparent match has provided major support to the notion that clumps are the products of disc fragmentation. As we have shown here, the match between the apparent typical clump mass and the Toomre mass in clumpy discs is primarily driven by the artificially inflated clump masses at low resolution and sensitivity (Figure \ref{fig:MassSelVsBound}). Essentially, only the high mass tail of the clump mass distribution is comparable to the Toomre mass, which is rather an upper limit for the characteristic clump mass set by fragmentation in the nonlinear regime, as explained in \citet{Tamburello2015}. The characteristic mass is much better matched with the modified Toomre mass proposed by \citet{Tamburello2015}, while the tail of very massive clumps that lie even above the Toomre mass (see Figure \ref{fig:MassSelVsBound}) results from later evolution, especially clump-clump mergers \citep{Tamburello2015}. 

While we showed that the observed clump properties can be reproduced using simulations in which clumps are primarily form through fragmentation processes, it does not imply that the fragmentation is the only reasonable scenario for most of the observed star forming clumps at high redshifts. Indeed, other processes such as minor mergers or accretion of the cores of disrupted satellites can also be plausible physical processes that lead to the formation of clumps, especially at the high mass tail of their distribution \citep{Mandelker2016}. If minor mergers are the dominant root for clump formation, their intrinsic size and masses are expected to be larger than what is expected from the disc fragmentation scenario, and one would expect the issues addressed in this paper to be less relevant and the intrinsically giant clumps remain large and massive even when increased resolution and sensitivity is achieved. The next generation of observations, especially those studying directly the gas phases with high resolution e.g., using the Atacama Large Millimeter Array (ALMA), should be able to uncover this alternative population of clumps.

We conclude that the discrepancies between clump properties found in observational studies, and between observations and simulations, can be significantly reduced by considering the different spatial resolutions accessible for the different studies, while still remaining in the context of the disc fragmentation scenario as the main origin for the clumps. While other important factors, such as the specific nature of feedback can drastically change the intrinsic clump properties \citep{Mayer2016}, testing model predictions against observations requires to align them with the proper resolution and sensitivity.

\section*{Acknowledgements}
This work is supported by the STARFORM Sinergia Project funded by the Swiss National Science Foundation.

%
%

\scalefont{0.94}
\setlength{\bibhang}{1.6em}
\setlength\labelwidth{0.0em}
\bibliography{Paper2}
\bibliographystyle{mnras}
\normalsize

%
%

\end{document}